\documentclass{llncs}

\usepackage{amssymb}
\usepackage{epsfig}

\newcommand{\ignore}[1]{}
\newcommand{\boxtheorem}{\hfill $\Box$}
\newcommand{\nit}[1]{{\it #1}}

\newcommand{\peer}[1]{\texttt{#1}}

\newcommand{\IC}{{\it IC}}
\newcommand{\trust}{{\it trust}}

\newcommand{\same}{{\it same}}
\newcommand{\less}{{\it less}}

\newcommand{\SiP}{\Sigma(\peer{P})}
\newcommand{\n}{~{\it not}~}
\newcommand{\aux}{{\it aux}}
\newcommand{\la}{\leftarrow}
\newcommand{\lp}{${\cal L}(\peer{P})$}
\newcommand{\icp}{\IC(\peer{P})}

\title{\bf \vspace*{-1cm}Query Answering in Peer-to-Peer Data Exchange Systems}

\vspace{-0.5cm}\author{{\bf  Leopoldo Bertossi} \and {\bf Loreto Bravo}\\
Carleton University, School of Computer Science,
Ottawa, Canada.\\
\{bertossi,lbravo\}@scs.carleton.ca}

\date{}
\institute{}
\begin{document}

 \maketitle

\begin{abstract} \vspace{-0.5cm}The problem of answering queries posed to a peer
who is a member of a peer-to-peer data exchange system is studied.
The answers have to be consistent wrt to both the local semantic
constraints and the data exchange constraints with other peers;
and must also respect certain trust relationships between peers. A
semantics for {\em peer consistent answers} under exchange
constraints and trust relationships is introduced and some
techniques for obtaining those answers are presented.
\end{abstract}

\vspace{-0.8cm}
\section{Introduction}\label{sec:intro}

\vspace{-3mm} In this paper the problem of answering queries posed
to a peer who is a member of a peer-to-peer data exchange system
is investigated. When a peer \peer{P} receives a query and is
going to answer it, it may need to consider both its own data and
the data stored at other peers' sites if those other peers are
related to \peer{P} by data exchange constraints (DECs). Keeping
the exchange constraints satisfied, may imply for peer \peer{P} to
get data from other peers to complement its own data, but also not
to use part of its own data. In which direction \peer{P} goes
depends not only on the exchange constraints, but also on the {\em
trust relationships} that \peer{P} has with other peers. For
example, if \peer{P} trust another peer \peer{Q}'s data more than
its own, \peer{P} will accommodate its data to \peer{Q}'s data in
order to keep the exchange constraints satisfied. Another element
to take into account in this process is a possible set of local
semantic constraints that each individual peer may have.

Given a network of peers, each with its own data,  and a
particular peer \peer{P} in it, a {\em solution for} \peer{P} is
-loosely speaking- a global database instance that respects the
exchange constraints and trust relationships \peer{P} has with its
immediate neighbors and stays as close as possible to the
available data in the system. Since the answers from \peer{P} have
to be consistent wrt to both the local semantic constraints and
the data exchange constraints with other peers, the {\em peer
consistent answers} (PCAs) from \peer{P} are defined as those
answers that can be retrieved from \peer{P}'s portion of data in
{\em every} possible solution for \peer{P}. This definition may
suggest that  \peer{P} may change other peers' data, specially of
those he considers less reliable, but this is not the case. The
notion of solution is used as an auxiliary notion to characterize
the correct answers from \peer{P}'s point of view. Ideally,
\peer{P} should be able to obtain its peer consistent answers just
by querying the already available local instances. This resembles
the approach to {\em consistent query answering} (CQA) in
databases \cite{pods99,bookChapter}, where answers to queries that
are consistent with given ICs are computed without changing the
original database.

We give a  precise semantics for peer consistent answers to
first-order queries. First for the {\em direct case}, where
transitive relationships between peers via ECs are not
automatically considered; and at the end, the {\em transitive
case}. We also illustrate by means of extended and representative
examples, mechanisms for obtaining PCAs (a full treatment is left
for an extended version of this paper). One of the approaches is
first order (FO) query rewriting, where the original query is
transformed into a new query, whose standard answers are the PCAs
to the original one. This methodology has intrinsic limitations.
The second, more general, approach is based on a specification of
the solutions for a peer as the stable models of a logic program,
which captures the different ways  the system stabilizes after
making the DECs and the trust relationships to be satisfied.

 We first recall the definition of database
repair that is used to characterize the consistent answers to
queries in single relational databases wrt certain integrity
constraints (ICs) \cite{pods99}. Given a relational database
instance $r$ with schema ${\cal R}$ (which includes a domain $D$),
$\Sigma(r)$ is the set
 of ground atomic formulas  $\{P(\bar{a}) ~|~ P \in {\cal R} \mbox{ and }  r \models
 P(\bar{a})\}$.

\begin{definition}\label{def:dist} \em \cite{pods99} ~(a) Let $r_1, r_2$ be
 database instances over  ${\cal R}$.
  The {\it distance}, $\Delta(r_1,r_2)$,
 between  $r_1$ and $r_2$ is the symmetric
 difference ~$\Delta(r_1,r_2)= (\Sigma(r_1) \setminus \Sigma(r_2)) \cup (\Sigma(r_2)
 \setminus \Sigma(r_1)).$\\
~ (b) For database instances $r,r_1, r_2$, we define
 $r_1 \leq_r r_2$ if $\Delta(r,r_1) \subseteq \Delta(r, r_2)$.\\
 (c) Let
 $\IC$ be a set of  ICs on ${\cal R}$.
 A {\em repair} of an instance $r$ wrt $\IC$ is a $\leq_r$-minimal
 instance $r'$, such that ~$r' \models \IC$.
 \boxtheorem
 \end{definition}
\vspace{-2mm}A repair of an instance $r$ is a consistent instance
 that minimally differs from $r$.

\vspace{-3mm}
\section{A Framework for P2P Data Exchange}

\vspace{-2mm} In this section we will describe the framework we
will use to formalize and address the problem of query answering
in P2P  systems.

\begin{definition}\em
A {\it $P\!2P$ data exchange system} $\frak P$ consists of:\\
(a) A finite set $\cal P$ of peers, denoted by \peer{A}, \peer{B},
...\\
(b) For each peer \peer{P}, a database schema ${\cal
R}(\peer{P})$, that includes a domain $D(\peer{P})$, and relations
$R(\peer{P}), ...$. However, it may be natural and convenient to
assume that all peers share a common, fixed, possibly infinite
domain, $D$. Each ${\cal R}(\peer{P})$ determines a FO language
${\cal L}(\peer{P})$. We assume that the schemata ${\cal
R}(\peer{P})$ are disjoint, being the domains the only possible
exception. ${\cal R}$ denotes the union of
the ${\cal R}(\peer{P})$s.\\
(c) For each peer \peer{P}, a database instance $r(\peer{P})$
corresponding to schema ${\cal R}(\peer{P})$.\\
(d) For each peer \peer{P}, a set of \lp-sentences $\IC(\peer{P})$
of ICs on
${\cal R}(\peer{P})$.\\
(e) For each peer \peer{P}, a collection $\Sigma({\peer{P}})$ of
{\em data exchange constraints} $\Sigma(\peer{P},\peer{Q})$
consisting of sentences written in the FO language for the
signature ${\cal R}(\peer{P}) \cup {\cal R}(\peer{Q})$, and the
$\peer{Q}$'s are (some of the) other peers
in ${\cal P}$.\\
(f) A {\em trust relation} $\trust \subseteq {\cal P} \times
\{\less, \same\} \times {\cal P}$, with the intended semantics
that when $(\peer{A}, \less, \peer{B}) \in \trust$, peer
$\peer{A}$ trusts itself less than $\peer{B}$; while $(\peer{A},
\same, \peer{B}) \in \trust$ indicates that $\peer{A}$ trusts
itself the same as $\peer{B}$. In this relation, the second
argument functionally depends on the other two.
 \boxtheorem
\end{definition}
\vspace{-2mm}Each peer \peer{P} is responsible for the update and
maintenance of its instance wrt $\icp$, independently from other
peers. In particular, we assume $r(\peer{P}) \models
\IC(\peer{P})$.\footnote{It would not be difficult to extend this
scenario to one that allows local violations of ICs. Techniques as
those described  in \cite{bookChapter} could be used in this
direction.} Peers may submit queries to another peer in accordance
with the restrictions imposed by the DECs and using the other
peer's relations appearing in them.

\vspace{-2mm}\begin{definition}\em (a) We denote with
$\overline{{\cal R}}(\peer{P})$ the schema consisting of ${\cal
R}(\peer{P})$ extended with the other peers' schemas that contain
predicates appearing in $\SiP$.\\
(b) For a peer \peer{P} and an instance $r$ on ${\cal
R}(\peer{P})$, we denote by $\bar{r}$, the database instance on
$\overline{{\cal R}}(\peer{P})$, consisting of the union of $r$
with all the peers' instances whose schemas appear in
$\overline{{\cal R}}(\peer{P})$.\\
(c) If $r$ is an instance over a certain schema ${\cal S}$ and
${\cal S}'$ is a subschema of ${\cal S}$, then $r|{\cal S}'$
denotes the restriction of $r$ to ${\cal S}'$. In particular, if
${\cal R}(\peer{P}) \subseteq {\cal S}$, then
$r|\peer{P}$ denotes the restriction of $r$ to ${\cal R}(\peer{P})$.\\
(d) We denote by ${\cal R}(\peer{P})^\less$ the union of all
schemata ${\cal R}(\peer{Q})$, with $(\peer{P},\less, \peer{Q})
\in \trust$. Analogously is ${\cal R}(\peer{P})^\same$ defined.
\boxtheorem
\end{definition}
\vspace{-1.5mm}From the perspective of a  peer \peer{P}, its own
database may be inconsistent wrt the data owned by another peer
\peer{Q} and the DECs in $\Sigma(\peer{P},\peer{Q})$.  Only when
\peer{P} trust \peer{Q} the same as or more than itself, it has to
consider \peer{Q}'s data. When \peer{P} queries its database,
these inconsistencies may have to be taken into account. Ideally,
the answers to the query obtained from \peer{P} should be
consistent with $\Sigma(\peer{P},\peer{Q})$ (and its own ICs
$\SiP$).
 In principle, \peer{P}, who is not allowed to change other peers' data,
  could try to repair its
database in order to satisfy  $\Sigma(\peer{P}) \cup \icp$. This
is not a realistic approach. Rather \peer{P} should solve its
 conflicts at query time, when it queries its own database and those of other peers.
Any answer obtained in this way should be sanctioned as correct
wrt to a precise semantics.

The semantics of peer consistent query answers for a peer \peer{P}
is given in terms of all possible minimal, virtual, simultaneous
repairs of the local databases that lead to a satisfaction of the
DECs while respecting \peer{P}'s trust relationships to other
peers. This repair process may lead to alternative global
databases called the {\em solutions}~ for \peer{P}. Next, the peer
consistent answers from \peer{P} are those that are  invariant wrt
to all its solutions. A peer's solution captures the idea that
only some peers' databases are relevant to \peer{P}, those whose
relations appear in its trusted exchange constraints, and are
trusted by \peer{P} at least as much as it trusts its own data. In
this sense, this is a ``local notion", because it does not take
into consideration transitive dependencies (but see Section
\ref{sec:beyond}).

\vspace{-2mm}\begin{definition} \label{def:solution} \em (direct
case) Given a peer \peer{P} in a P2P data exchange system $\frak
P$ and an instance $r$ on ${\cal R}$,  we say that an instance
$r'$ on ${\cal R}$ is a {\em solution for} \peer{P} if,
simultaneously:~ (a) $r' \models \Sigma(\peer{P}) \cup
\IC(\peer{P})$.~ (b) $r'|P = r|P$ for every predicate $P \notin
\overline{{\cal R}}(\peer{P})$.~ (c) There are instances $r_1,
r_2$ over ${\cal R}$ satisfying:~ (c1) $r_2 = r'$. ~(c2) $r_1$ is
a repair of $r$ wrt $\bigcup \{\Sigma(\peer{P},\peer{Q}) ~|~
(\peer{P},\less,\peer{Q}) \in \trust \}$, with $r_1|\peer{Q} =
r|\peer{Q}$ whenever $(\peer{P},\less,\peer{Q}) \in \trust$ or
$(\peer{P},\same,\peer{Q}) \in \trust$. ~(c3) $r_2$ is a repair of
$r_1$ wrt $\bigcup \{\Sigma(\peer{P},\peer{Q}) ~|~
(\peer{P},\same,\peer{Q}) \in \trust\}$, such that $r_2 \models
\Sigma(\peer{P},\peer{Q})$ and $r_2|\peer{Q} = r_1|\peer{Q}$ for
those peers $\peer{Q}$ with $(\peer{P},\less,\peer{Q}) \in
\trust$. \boxtheorem
\end{definition}
\vspace{-1mm}The  {\em solutions} for a peer  are used as a
conceptual, auxiliary tool to characterize the semantically
correct answers to a peer's queries. We are not interested in
computing a peer's solutions {\em per se}. Solutions (and repairs)
are virtual and may be only partially computed if necessary, if
this helps us to compute the correct answers obtained in/from a
peer. The ``changes" that are implicit in the definition of
solution via the set differences  are expected to be minimal wrt
to sets of  tuples which are inserted/deleted into/from the
tables.

 In intuitive terms, a solution for $\peer{P}$ repairs the global
instance, but leaves unchanged the tables that do not appear in
its trusted ICs and those tables that  belong to peers that are
more trusted by him than himself. With this condition, $\peer{P}$
first tries to change its own tables according to what the
dependencies to more trusted peers of peers prescribe. Next,
keeping those more trusted dependencies satisfied, it tries to
repair its or other peers' data, but only considering those peers
who are equality trusted as itself.

In these definitions we find clear similarities with the
characterization of consistent query answers in
 single relational databases \cite{bookChapter}.
However, in P2P query answering, repairs may involve data
associated to different peers, and also a notion of priority that
is related to the trust relation (other important differences are
discussed below).

\vspace{-2mm}
\begin{example} \label{ex:peer} Consider a P2P data system with peers
\peer{P1}, \peer{P2}, \peer{P3}, and schemas ${\cal R}_i = \{R^i,
\ldots \}$, and instances $r^i$, $i = 1,2,3$, resp.; and:~  (a)
$r^1 = \{R^1(a,b), R^1(s,$ $t)\}$, ~~$r^2 = \{R^2(c,d),
R^2(a,e)\}$, ~~$r^3 = \{R^3(a,f), R^3(s,u)\}$.~ (b) $\trust = \{
~(\peer{P1},\less,$ $ \peer{P2}), ~~(\peer{P1},\same,
\peer{P3})~\}$.~ (c) $\Sigma(\peer{P1},\peer{P2}) = \{~ \forall x
y(R^2(x,y) \rightarrow R^1(x,y))~\}$;
~$\Sigma(\peer{P1},\peer{P3}) $ $= \{~\forall x y z(R^1(x,y)
\wedge R^3(x,z) ~~\rightarrow~~ y = z)~\}$.

Here, the global instance is $r = \{R^1(a,b), R^1(s,t), R^2(c,d),
R^2(a,e),$ $ R^3(a,f),$ $ R^3(s,u)\}$. The solutions for \peer{P1}
are obtained by first repairing $r$ wrt the relationship between
\peer{P1} and \peer{P2}. Then $r_1$ in condition (c2) in
Definition \ref{def:solution} is $r_1 = \{R^1(a,b), R^1(s,t),$ $
R^1(c,d), R^1(a,e),$ $ R^2(c,d), R^2(a,e), $ $ R^3(a,f),
R^3(s,u)\}$. In this example there is only one repair at this
stage, but in other situations there might be several. Now, this
repair has to be repaired in its turn wrt the data dependency
between \peer{P1} and \peer{P3} (but keeping the relationship
between \peer{P1} and \peer{P2} satisfied). In this case, we
obtain only two repairs, $r' = \{R^1(a,b), R^1(s,t),  R^1(c,d),$ $
R^1(a,e), R^2(c,d), R^2(a,e)\}$; and $r'' = $ $\{$ $R^1(a,b),
R^1(c,d), R^1(a,e), R^2(c,d),$ $R^2(a,$ $e), R^3(s,u)\}$. These
are the only solutions for peer \peer{P1}. \boxtheorem
\end{example}
\vspace{-2mm}The  minimization involved in  a solution is similar
to a prioritized minimization (with some predicates that are kept
fixed) found in non-monotonic reasoning \cite{Lifschitz87}.
Actually, the notion of consistent query answer -even the one
based on the non prioritized version of repair (c.f. Definition
\ref{def:dist})- is a non-monotonic notion
\cite{bookChapter}.\footnote{A circumscriptive approach to
database repairs was given in \cite{amai}. It should not be
difficult to extend that characterization to capture the peer
solutions.}

Notice that the notion of a solution for a peer \peer{P} is a
``local notion" in the sense that it considers the ``direct
neighbors" of \peer{P} only. One reason for considering this case
is that \peer{P} does not see beyond its neighbors; and when
\peer{P} requests data to a neighbor, say \peer{Q}, the latter may
decide -or even \peer{P} may decide a priori and in a uniform way-
that for \peer{P} it is good enough to accommodate its data to its
neighbors alone, without considering any transitive dependencies.
In section \ref{sec:beyond} we will explore the case of
interrelated dependencies.

 Now we can define which are
the intended answers to a query posed to a peer, from the
perspective of that peer.

\begin{definition} \label{def:peercons} \em
Given a FO query $Q(\bar{x}) \in {\cal L}(\peer{P})$, posed to
peer \peer{P}, a ground tuple $\bar{t}$ is {\em peer consistent}
for \peer{P} iff $r'|\peer{P} \models Q(\bar{t})$ for every
solution $r'$ for \peer{P}. \boxtheorem
\end{definition}
Notice that this definition is relative to a fixed peer, and not
only because the query is posed to one peer and in its query
language, but also because this notion is based on the direct
notion of solution for a single peer.

Peer consistent answers to queries can be obtained by using
techniques similar to those developed for CQA, for example, query
rewriting based techniques \cite{pods99,bookChapter}. However,
there are important differences, because now we have some fixed
predicates in the repair process.

\vspace{-2mm}
\begin{example} \label{ex:peer2} (example \ref{ex:peer} continued) If \peer{P1} is
posed the query $Q\!: R^1(x,y)$ asking for the tuples in relation
$R^1$, we first rewrite the query by considering the exchange
dependencies in $\Sigma(\peer{P1},\peer{P2})$,  obtaining~ $Q'\!:
R^1(x,y) \vee R^2(x,y),$ which basically has the effect of
bringing \peer{P2}'s data into \peer{P1}. Next, the exchange
dependency $\Sigma(\peer{P1},\peer{P3})$ is considered, and now
the query  is rewritten into
\vspace{-3mm}\begin{equation}\label{eq:rew2} Q''\!: [R^1(x,y)
\wedge \forall z_1(R^3(x,z_1) \wedge \neg \exists z_2 R^2(x,z_2)
~\rightarrow~ z_1 = y)] ~~\vee~~ R^2(x,y).\vspace{-3mm}
\end{equation}
In order to answer this query, \peer{P1} will first issue a query
to \peer{P2} to retrieve the tuples in $R^2$; next, a query is
issued to \peer{P3} to leave outside  $R^1$ those tuples  that
appear with the same first but not the same second argument in
$R^3$, as long as the conflicting tuple in $R^2$ is ``protected"
by a tuple in $R^3$ which has the same key as a the two
conflicting tuples in $R^1$ and $R^3$ ($R^1(a,b)$ above).  The
answers to query (\ref{eq:rew2}) are $(a,b), (c,d), (a,e)$,
precisely the peer consistent answers to query $Q$ for peer
\peer{P1} according to their semantic definition. \boxtheorem
\end{example}
\vspace{-2mm}Notice that a  query $Q$ may have peer consistent
answers for a peer which are not answers to $Q$ when the peer is
considered in isolation. This makes sense, because the peer may
import data from other peers. This is another difference with CQA,
where all consistent answers are answers to the original
query\footnote{At least if the ICs are {\em generic}
\cite{bookChapter}, i.e. they do not imply by themselves the
presence/absence of any particular ground tuple in/from the
database.}.

The query rewriting approach suggested in Example \ref{ex:peer2}
differs from the one used for CQA. In the latter case, literals in
the query are resolved (using {\em resolution}) against the ICs in
order to generate residues that are appended as extra conditions
to the query, in an iterative process. In the case of P2P data
systems, the query may have to be modified in order to include new
data that is located at a different peer's site. This cannot be
achieved by imposing extra conditions alone -as in the query
rewriting based consistent query answering- but instead, by
relaxing the query in some sense.

Instead of pursuing and fully developing a FO query transformation
approach to query answering in P2P systems, we will propose (see
Section \ref{sec:anssets}) an alternative methodology based on
answer set programming, which is more general. Furthermore, since
query answering in P2P systems already includes some sufficiently
complex cases of CQA, a FO query rewriting approach to P2P query
answering is bound to have important limitations in terms of
completeness, as in CQA \cite{bookChapter}; for example in the
case of existential queries and/or existential DECs.

\vspace{-3mm}
\section{Referential Exchange Constraints}
\label{sec:anssets}

\vspace{-2mm} In most applications we may expect the exchange
constraints to be inclusion dependencies or referential
constraints, i.e. formulas of the form
\vspace{-3mm}\begin{equation}\label{eq:incl} \forall \bar{x}
\exists \bar{y} (R^\peer{Q}(\bar{x}) \wedge \cdots ~~ \rightarrow
R^\peer{P}(\bar{z},\bar{y}) \wedge \cdots), \vspace{-2mm}
\end{equation}
where $R^\peer{Q}, R^\peer{P}$ are relations for peers \peer{Q}
and \peer{P}, resp., the dots indicate some possible additional
conditions, most likely expressed in terms of built-ins, $\bar{z}
\subseteq \bar{x}$ (if $\bar{y} = \emptyset$ and $\bar{z} =
\bar{x}$, and no additional conditions are given, we have a full
inclusion dependency, like $\Sigma(\peer{P1},\peer{P2})$ in
Example \ref{ex:peer}).

An exchange constraint of the form (\ref{eq:incl}) will most
likely belong to $\Sigma(\peer{P},\peer{Q})$, i.e. to peer
\peer{P}, who wants to import data from the more trustable peer
\peer{Q}. It could also belong to $\peer{Q}$, if this peer wants
to validate its own data against the data at \peer{P}'s site.
Section \ref{sec:mixed} shows an example of a more involved
referential constraint.

An answer set programming approach to the specification of
solutions for a peer can be developed. In spirit, those
specifications would be similar to those of repairs of single
relational databases under referential integrity constraints
\cite{semantics03}. However -as already seen in Examples
\ref{ex:peer} and \ref{ex:peer2}- there are important differences
with CQA. In Section \ref{sec:mixed} we give an example that shows
the main issues around this kind specification.\footnote{A
detailed and complete approach will be found in an extended
version of this paper.}

\vspace{-3mm}
\subsection{An extended example} \label{sec:mixed}

\vspace{-2mm} Consider a P2P data exchange system  with peers
\peer{P} and \peer{Q}, with schemas $\{R_1(\cdot,$
$\cdot),R_2(\cdot,\cdot)\}$,  $\{S_1(\cdot,\cdot),$
$S_2(\cdot,\cdot)\}$, resp. Peer \peer{P} also has the exchange
constraint \vspace{-2mm}\begin{equation}\label{eq:mixed} \forall x
\forall y \forall z \exists w (R_1(x,y) \wedge S_1(z,y)
~\rightarrow~ R_2(x,w) \land S_2(z,w)), \vspace{-2mm}
\end{equation}
which mixes tables of the two peers  on each side of the
implication.

Let us assume that peer \peer{P} is querying his database, but
subject to its DEC (\ref{eq:mixed}). We will consider the case
where $(\peer{P},\less,\peer{Q}) \in \trust$, i.e. \peer{P}
considers \peer{Q}'s data more reliable than his own. If
(\ref{eq:mixed}) is satisfied by the combination of the data in
\peer{P} and \peer{Q}, then the current global instance
constitutes \peer{P}'s solution. Otherwise, alternative solutions
for \peer{P} have to be found, keeping \peer{Q}'s data fixed in
the process. This is the case, where there are ground tuples
$R_1(d,m) \in r(\peer{P}), S_2(a,m) \in r(\peer{Q})$, such that
for no $t$ it  holds both $R_2(d,t) \in r(\peer{P})$ and $S_2(a,t)
\in r(\peer{Q})$.

Obtaining peer consistent answers to queries for peer \peer{P}
amounts to virtually restoring the satisfaction of
(\ref{eq:mixed}), actually by  virtually modifying \peer{P}'s
data. In order to specify \peer{P}'s modified relations, we
introduce virtual versions $R_1', R_2'$ of $R_1, R_2$, which will
contain the data in peer \peer{P}'s solutions. In consequence, at
the solution level, we have the relations $R_1', R_2', S_1, S_2$.
 Since \peer{P}
is querying its database, its original queries will be expressed
in terms of relations $R_1', R_2'$ only (plus possible built-ins).

The contents of the virtual relations $R'_1, R'_2$ will be
obtained from the contents of the material sources $R_1, R_2, S_1,
S_2$.\footnote{We can observe that the virtual relations can be
seen as virtual global relations in a virtual data integration
system \cite{levy00,lenzerini02}. For a more detailed comparison
between data integration and peer data management systems see
\cite{halevyICDE03,tatarinov03}.} Since $S_1, S_2$ are fixed, the
satisfaction of (\ref{eq:mixed}) requires  $R_1'$ to be a subset
of $R_1$, and $R_2'$, a superset of $R_2$. The specification of
these relations can be done in disjunctive extended logic programs
with answer set (stable model) semantics \cite{glELPb}. The first
 rules for the specification program $\Pi$ are:
\vspace{-5mm}\begin{eqnarray}
R_1'(x,y) &\leftarrow& R_1(x,y), \n \neg R_1'(x,y) \label{eq:cwa1}\\
 R_2'(x,y)
&\leftarrow& R_2(x,y), \n \neg
R_2'(x,y),\label{eq:cwa2}\vspace{-5mm}
\end{eqnarray}
which specify that, by default, the tuples in the source relations
are copied into the new virtual versions,  but with the exception
of  those that may have to be removed in order to satisfy
(\ref{eq:mixed}) (with $R_1,R_2$ replaced by $R'_1,R'_2$). Some of
the exceptions for $R'_1$ are specified by
\vspace{-2mm}\begin{eqnarray} \neg R_1'(x,y) &\leftarrow&
R_1(x,y), S_1(z,y), \n \aux_1(x,z), \n
aux_2(z)\label{eq:rep1}\\ \aux_1(x,z) &\la& R_2(x,w), S_2(z,w) \label{eq:rep2}\\
\aux_2(z) &\la& S_2(z,w) \label{eq:rep3}.
\end{eqnarray}
That is, $R_1(x,y)$ is deleted if it participates in a violation
of (\ref{eq:mixed}) (what is captured by the first three literals
in the body of (\ref{eq:rep1}) plus rule (\ref{eq:rep2})), and
there is no way to restore consistency by inserting a tuple into
$R_2$, because there is no possible matching tuple in $S_2$ for
the possibly new tuple in $R_2$ (what is captured by the last
literal in the body of (\ref{eq:rep1}) plus rule (\ref{eq:rep3})).
In case there is such a tuple in $S_2$, then we have the
alternative of either deleting a tuple from $R_1$ or inserting a
tuple into $R_2$: \vspace{-3mm}\begin{eqnarray} \neg R_1'(x,y)
\vee R_2'(x,w)
&\la& R_1(x,y), S_1(z,y), \n \aux_1(x,z), S_2(z,w), \nonumber \\
&& {\it choice}((x,z),w). \vspace{-10mm}\label{eq:disj}
\end{eqnarray}  That is, in
case of a violation of  (\ref{eq:mixed}), when there is tuple of
the form $(a,t)$ in $S_2$ for the combination of values $(d,a)$,
then the {\em choice operator} \cite{GPSZ91} non deterministically
chooses a unique value for $t$, so that the tuple $(d,t)$ is
inserted into $R_2$ as an alternative to deleting $(d,m)$ from
$R_1$. Notice that no exceptions are specified for $R'_2$, what
makes sense since $R'_2$ is a superset of $R_2$. In consequence,
the negative literal in the body of (\ref{eq:cwa2}) can be
eliminated. However, new tuples can be inserted into $R'_2$, what
is captured by rule (\ref{eq:disj}). Finally, the program must
contain as facts the tuples in the original relations
$R_1,R_2,S_1,S_2$.

In the case where \peer{P} equally trusts himself and \peer{Q},
both \peer{P}  and \peer{Q}s' relations become flexible when
searching for a solution for \peer{P}. The program becomes more
involved, because now $S_1, S_2$ may also change. In consequence,
virtual versions for them should be introduced and specified.

\vspace{-3mm}
\subsection{Considerations on specifications of peers' solutions}

\vspace{-2mm} The  example we presented in Section \ref{sec:mixed}
shows the main issues in the  specification of a peer's solutions
under referential exchange constraints. If desired, the choice
operator can be replaced by a predicate that can be defined by
means of extra rules, producing the so-called stable version of
the choice program \cite{GPSZ91}. This stable version has a
completely standard answer set semantics.

The peer's solutions are in one to one correspondence with the
answer sets of the program. In the previous example, each solution
$r^S$ for peer \peer{P} coincides with the original, material,
global instance for the tables other than $R_1, R_2$, whereas the
contents $r_1^S, r_2^S$ for these two are of the form $r_i^S =
\{\bar{t}~|~ R'_i(\bar{t}) \in S\}$, where $S$ is an answer set of
program $\Pi$. The absence of solutions for a peer will thus be
captured by the non existence of answer sets for program $\Pi$.

Program $\Pi$ represents in a compact form all the solutions for a
peer; in consequence, the peer consistent answers to a query posed
to the peer  can be obtained by running the query, expressed as a
query program in terms of the virtually repaired tables, in
combination with the specification program $\Pi$. The answers so
obtained will be those that hold for all the possible solutions if
the program is run under the skeptical answer set semantics. As
for consistent query answering, a system like DLV \cite{dlvSys}
can be used for this purpose.

For example, the query $Q\!(x,z): ~\exists y (R_1(x,y) \wedge
R_2(z,y))$ issued to peer \peer{P}, would be peer consistently
answered by running the query program $Ans_Q(x,z) \la R'_1(x,y),
R'_2(x,y)$ together with program $\Pi$. Although only (the new
versions of) \peer{P}'s relations appear in the query, the program
may make \peer{P} import data from \peer{Q}.

If a peer has local ICs that have to be satisfied and a program
has been used to specify its solutions, then the program should
take care of those constraints. One simple way of doing this
consists in using program denial constraints. If in Section
\ref{sec:mixed} we had for peer \peer{P} the local functional
dependency (FD) $\forall x \forall y \forall z(R_1(x,y) \wedge
R_1(x,z) \rightarrow y = z)$, then program would include the
program constraint ~$\la R_1(x,y), R(x,z), y \neq z$, which would
have the effect of pruning those solutions (or models of the
program) that do not satisfy the FD. DLV, for example, can handle
program denial constraints \cite{leoneCorr}.

A more flexible alternative to keeping the local ICs satisfied,
consists in having the specification program split in two layers,
where the first one builds the solutions, without considering the
local ICs, and the second one, repairs the solutions wrt the local
ICs, as done with single inconsistent relational databases
\cite{semantics03}.

Finally, we should notice that obtaining peer consistent answers
has at least the data complexity of consistent query answering,
for which some results are known
\cite{chomickiCorr,fuxman,tanosPods03}. In the latter case, for
common database queries and ICs, $\Pi^P_2$-completeness is easily
achieved. On the other side, the problem  of skeptical query
evaluation from the disjunctive programs we are using for P2P
systems is also $\Pi^P_2$-complete in data complexity
\cite{complexLP}. In this sense, the logic programs are not
contributing with additional complexity to our problem.

\vspace{-4mm}
\section{Discussion and Extensions}\label{sec:conclusions}

\vspace{-2mm}
\subsection{Optimizations}
\vspace{-2mm} It is possible to perform some optimizations on the
program, to make its evaluation simpler. Disjunctive program under
the stable model semantics are more complex than non disjunctive
programs \cite{complexLP}. However, it is known that a disjunctive
program can be transformed into a non disjunctive program if the
program is head-cycle free (HCF) \cite{rachel,infComp}.
Intuitively speaking, a disjunctive program is HCF if there  are
no cycles involving two literals in the head of a same rule, where
a link is established from a literal to another if the former
appears positive in the body of a rule, and the latter appears in
the head of the same rule. These considerations about HCF programs
hold for programs that do not contain the choice operator, i.e.
they might not automatically apply to our programs that specifies
the solutions for a peer under referential constraints. However,
it is possible to prove that a disjunctive choice program $\Pi$ is
HCF when the program obtained from $\Pi$ by removing its choice
goals is HCF. \cite{submitted}.

\vspace{-2mm}\begin{example}\label{ex:mixed2} Consider the choice
program $\Pi$ presented in Section \ref{sec:mixed}. If the choice
operator is eliminated from rule (\ref{eq:disj}), we are left with
the rule \vspace{-2mm}
$$\neg R_1'(x,y) \vee R_2'(x,w) \la R_1(x,y), S_1(z,y), \n
\aux_1(x,z), S_2(z,w).$$ The resulting program is HCF and then
rule (\ref{eq:disj}) can be replaced by  two rules: \vspace{-2mm}
\begin{eqnarray*} \neg R_1'(x,y) &\la& R_1(x,y),
S_1(z,y), \n \aux_1(x,z), S_2(z,w), \n R_2'(x,w),
\nonumber \\ &&  {\it choice}((x,z),w).\\
R_2'(x,w) &\la& R_1(x,y), S_1(z,y), \n \aux_1(x,z), S_2(z,w), \n
\neg R_1'(x,y), \nonumber \\ &&  {\it choice}((x,z),w).
\hspace{6.9cm} \Box
\end{eqnarray*}
\end{example}

\vspace{-8mm}
\subsection{A LAV approach}\label{sec:lav}

\vspace{-2mm} The logic programming-based approach proposed in
Section \ref{sec:mixed} can be seen assimilated to  the
global-as-view (GAV) approach to virtual data integration
\cite{lenzerini02}, in the sense that the tables in the solutions
are specified as views over the peer's schemas. However, a
local-as-view (LAV) approach could also be attempted. In this
case, we also introduce virtual, global versions of $S_1, S_2$.
The relations in the sources have to be defined as views of the
virtual relations in a solution, actually, through the following
specification of a virtual integration system \cite{GM99}

\begin{center}
 \begin{tabular}{|l| c| c|}\hline
 View definitions & label & source\\ \hline \hline
 $R_1(x,y) \leftarrow R_1'(x,y)$&  {\it closed} & $r_1$\\ \hline
 $R_2(x,y) \leftarrow R_2'(x,y)$ & {\it open} & $r_2$\\ \hline
 $S_1(x,y) \leftarrow S_1'(x,y)$ & {\it clopen} & $s_1$\\ \hline
 $S_2(x,y) \leftarrow S_2'(x,y)$ & {\it clopen} & $s_2$\\
 \hline
\end{tabular}
\end{center}

\noindent Here the $r_i, s_j$ are the original material extensions
of relations $R_i,S_j$. The labels for the sources are assigned on
the basis of the view definitions in the first column, the IC
(\ref{eq:mixed}) and the trust relationships; in the latter case,
by the fact that $R_1, R_2$ can change, but not $S_1, S_2$. More
precisely, the label in the first row corresponds to the fact that
(\ref{eq:mixed}) can be satisfied by deleting tuples from $R_1$,
then the contents of the view defined in there must be contained
in the original relation $r_1$ (the material source). The label in
the second row indicates that we can insert tuples into $R_2$ to
satisfy the constraint, and then, the extension of the solution
contains the original source $r_2$. Since, $S_1, S_2$ do not
change, we declare them as both closed and open, i.e. clopen.

If a query is posed to, say peer \peer{P}, it has to be first
formulated in terms of $R'_1, R'_2$, and then it can be peer
consistently answered by querying the integration system subject
to the global IC:~
$\forall x \forall y \forall z \exists w (R'_1(x,y) \wedge
S'_1(z,y) \rightarrow R'_2(x,w) \land S'_2(z,w)).$ 
A methodology that is similar to the one applied for consistently
querying virtual data integration systems under  LAV can be used.
In \cite{fqas02,IJCAI03} methodologies for open sources are
presented, and in \cite{chapDag03} the mixed case with both open,
closed and clopen sources is treated. However, there are
differences in our P2P scenario; and those methodologies need to
be adjusted.

The methodology presented in \cite{chapDag03} for CQA in virtual
data integration is based on a three-layered answer set
programming specification of the repairs of the system: a first
layer specifies the contents of the global relations in the
minimal legal instances (to this layer only open and clopen
sources contribute), a second layer consisting of program denial
constraints that prunes the models that violate the closure
condition for the closed sources; and a third layer specifying the
minimal repairs of the legal instances \cite{fqas02} left by the
other layers wrt the global ICs. For CQA, repairs are allowed to
violate the original labels.

In our P2P scenario, we want, first of all, to consider only the
legal instances that satisfy the mapping in the table and that, in
the case of closed sources include the maximum amount of tuples
from the sources (the virtual relations must be kept as close as
possible to their original, material versions). For the kind of
mappings that we have in the table, this can be achieved by using
exactly the same kind of specifications presented in
 in \cite{chapDag03} for the mixed case, {\em but} considering the closed sources
as clopen. In doing so, they will contribute to the program with
both rules that import their contents into the system (maximizing
the set of tuples in the global relation) and denial program
constraints. Now, the trust relation also makes a difference. In
order for the virtual relations to satisfy the original labels,
that in their turn capture the trust relationships, the rules that
repair the chosen legal instances will  consider only tuple
deletions (insertions) for  the virtual global relations
corresponding to the closed (resp. open) sources. For clopen
sources the rules can neither add nor delete tuples.\footnote{This
preference criterion for a subclass of the repairs is similar to
the {\em loosely-sound semantic} for integration of open sources
under GAV \cite{lembo}.} This methodology can handle universal and
simple referential DECs  (no cycles and single atom consequents,
conditions that are imposed by the repair layer of the program),
which covers a broad class of DECs. The DEC in (\ref{eq:mixed})
does not fall in this class, but the repair layer can be easily
adjusted in order to generate the solutions for peer \peer{P}. Due
to space limitations, the program is given in the appendix.

\vspace{-3mm}
\subsection{Beyond direct solutions} \label{sec:beyond}

\vspace{-2mm} It is natural to considers {\em transitive data
exchange dependencies}. This is a situation that arises when, e.g.
a peer \peer{A}, that is being queried, gets data from another
peer \peer{B}, who in its turn -and without \peer{A} possibly
knowing- gets data from a third peer \peer{C} to answer \peer{A}'s
request. Most likely there won't be any explicit DEC from \peer{A}
to \peer{C} capturing this transitive exchange; and we do not want
to derive any.

In order to attack peer consistent query answering in this more
complex scenario, it becomes necessary to integrate the local
solutions, what can be achieved by integrating the ``local"
specification programs. In this case, it is much more natural and
simpler than extending the definition of solutions for the direct
case, to define the semantics of a peer's (global) solutions
directly as the answer sets of the combined programs. Of course,
there might be no solutions, what is reflected in the absence of
stable models for the program. A problematic case appears when
there are implicit cyclic dependencies \cite{halevyICDE03}.

\vspace{-1mm}
\begin{example}\label{ex:global} (example in Section \ref{sec:mixed} continued) Let
us consider another peer \peer{C}. The following exchange
constraint $\Sigma_{\peer{Q},\peer{C}}\!:~ \forall x\forall
y(U(x,y) \rightarrow S_1(x,y))$ exists from \peer{Q} to \peer{C}
and $(\peer{Q},\less,\peer{C}) \in \trust$, meaning that \peer{Q}
trusts \peer{C}'s data more than its own. When \peer{P} requests
data from \peer{Q}, the latter will  request data from \peer{C}'s
relation $U$. Now, consider the peer instances: $r_1=\{(a,b)\},
s_1=\{\},$ $r_2=\{\}$, $s_2=\{(c,e),(c,f)\}$ and $u=\{(c,b)\}$. If
we analyze each peer locally,  the solution for \peer{Q} would
contain the tuple $S_1(c,b)$ added; and  \peer{P} would have only
one solution, corresponding to the original instances, because the
DEC is satisfied without making any changes. When considering them
globally, the tuple that is locally added  into \peer{Q} requires
 tuples to be added and/or deleted into/from
\peer{P} in order to satisfy the DEC. The combined program that
specifies the global solutions consists of  rules (\ref{eq:cwa1}),
(\ref{eq:cwa2}),(\ref{eq:rep2}), (\ref{eq:rep3}) plus
((\ref{eq:new1}), (\ref{eq:new2}) replace (\ref{eq:rep1}),
(\ref{eq:disj}), resp.) \vspace{-2mm}\begin{eqnarray} \neg
R_1'(x,y) &\leftarrow&
R_1(x,y), S_1'(z,y), \n \aux_1(x,z), \n aux_2(z)~~~~~ \label{eq:new1}\\
\neg R_1'(x,y) \vee R_2'(x,w) &\la& R_1(x,y), S_1'(z,y), \n
\aux_1(x,z), S_2(z,w), \nonumber \\ && {\it choice}((x,z),w) \label{eq:new2}\\
S_1'(x,y)&\leftarrow& S_1(x,y), \n \neg S_1'(x,y)\\ S_1'(x,y)&\la&
U(x,y), \n S_1(x,y). \label{eq:GS1}
\end{eqnarray}
The solutions obtained from the stable models of the program are
precisely the expected ones: $r_1= \{S_2(c,e), S_2(c,f), U(c,b),
S_1'(c,b), R_2'(a,f), R_1'(a,b)\}$, $r_2 = \{S_2(c,e), S_2(c,f),
U(c,b), S_1'(c,b)\}$,  $r_3 = \{ S_2(c,e), S_2(c,f), U(c,b),
S_1'(c,b),$ $
 R_2'(a,e),  R_1'(a,b)\}$. \boxtheorem
\end{example}

\vspace{-1mm}\noindent   {\bf Acknowledgements:}~ Research funded
by NSERC Grant 250279-02. L. Bertossi is Faculty Fellow of the IBM
Center for Advanced Studies, Toronto Lab. We are grateful for
stimulating conversations with Ariel Fuxman, who, in particular,
proposed to consider the example in Section \ref{sec:mixed}.

\newpage \vspace*{-1.5cm}
\section{Appendix:}
\vspace{-3mm}The following answer set program specifies the
solutions for the example in Section \ref{sec:mixed} following a
LAV approach to P2P data exchange (see Section \ref{sec:lav}).
Assume that the peers have the following instances:~
$r_1=\{(a,b)\}, s_1=\{(c,b)\},$ $r_2=\{\}$ and
$s_2=\{(c,e),(c,f)\}$. Then, the facts of the program are: ~
$R_1(a,b),$ $S_1(c,b),
    S_2(c,e),  S_2(c,f)$. The layer that specifies the preferred
    legal instances contains the following rules:
\begin{eqnarray*}
    R_1'(X,Y,t_d) &\leftarrow& R_1(X,Y).\\
    S_1'(X,Y,t_d) &\leftarrow& S_1(X,Y).\\
    R_2'(X,Y,t_d) &\leftarrow& R_2(X,Y).\\
    S_2'(X,Y,t_d) &\leftarrow& S_2(X,Y).\\
    &\leftarrow& R_1'(X,Y,t_d), R_1(X,Y).\\
    &\leftarrow& S_1'(X,Y,t_d), S_1(X,Y).\\
    &\leftarrow& S_2'(X,Y,t_d), S_2(X,Y).
    \end{eqnarray*}
The layer that specifies the repairs of the legal instances
contains the following rules. The annotation constants in the
third arguments in the relations are used as auxiliary elements in
the repairs process \cite{semantics03}. The choice operator has
been unfolded, producing the stable version of the choice program.
\begin{eqnarray*}
    R_1'(X,Y,t_{ss}) &\leftarrow& R_1'(X,Y,t_d), \n R_1'(X,Y,f_a).\\
    R_1'(X,Y,t_{ss}) &\leftarrow& R_1'(X,Y,t_a).\\
    &\leftarrow&R_1'(X,Y,t_a), R_1'(X,Y,f_a).\\
    S_1'(X,Y,t_{ss}) &\leftarrow& S_1'(X,Y,t_d), \n S_1'(X,Y,f_a).\\
    S_1'(X,Y,t_{ss}) &\leftarrow& S_1'(X,Y,t_a).\\
    &\leftarrow&S_1'(X,Y,t_a), S_1'(X,Y,f_a).\\
    R_2'(X,Y,t_{ss}) &\leftarrow& R_2'(X,Y,t_d), \n R_2'(X,Y,f_a).\\
    R_2'(X,Y,t_{ss}) &\leftarrow& R_2'(X,Y,t_a).\\
    &\leftarrow&R_2'(X,Y,t_a), R_2'(X,Y,f_a).\\
    S_2'(X,Y,t_{ss}) &\leftarrow& S_2'(X,Y,t_d), \n S_2'(X,Y,f_a).\\
    S_2'(X,Y,t_{ss}) &\leftarrow& S_2'(X,Y,t_a).\\
    &\leftarrow&S_2'(X,Y,t_a), S_2'(X,Y,f_a).\\
    R_1'(X,X,f_a) &\leftarrow& R_1'(X,Y,t_d), S_1'(Z,Y,t_d), \n aux_1(X,Z), \\
    &&\n aux_2(Z).\\
    aux_1(X,Z) &\leftarrow& R_2'(X,U,t_d), S_2'(Z,U,t_d).\\
    aux_2(Z) &\leftarrow& S_2'(Z,W,t_d).\\
    R_1'(X,Y,f_a) \vee R_2'(X,W,t_a) &\leftarrow& R_1'(X,Y,t_d), S_1'(Z,Y,t_d), \n
    aux_1(X,Z),\\
                 &&              S_2'(Z,W,t_d), chosen(X,Z,W).\\
    chosen(X,Z,W) &\leftarrow& R_1'(X,Y,t_d), S_1'(Z,Y,t_d), \n
    aux_1(X,Z),\\
                   &&  S_2'(Z,W,t_d), \n \nit{diffchoice}(X,Z,W).\\
    \nit{diffchoice}(X,Z,W) &\leftarrow& chosen(X,Z,U), S_2'(Z,W,t_d),  U \neq W.\\
\end{eqnarray*}
The following are the stable models of the program:\\

\noindent
\begin{tabular}{lp{11cm}} $M_1$= &$\{R_1(a,b),$ $ S_1(c,b),$ $
S_2(c,e),$ $ S_2(c,f),$ $ R_1'(a,b,td),$ $ S_1'(c,b,td),$ $
S_2'(c,e,td),$ $ S_2'(c,f,td),$
 $ \aux_2(c),$ $ S_1'(c,b,tss),$ $ S_2'(c,e,tss),$ $
S_2'(c,f,tss),$ $ R_1'(a,b,tss),$ $ \nit{diffchoice}(a,c,e),$ $
chosen(a,c,f),$ $ R_2'(a,f,ta),$ $ R_2'(a,f,tss)\} $ \\
$M_2$=& $\{R_1(a,b),$ $ S_1(c,b),$ $ S_2(c,e),$ $ S_2(c,f),$ $
R_1'(a,b,td),$ $ S_1'(c,b,td),$ $ S_2'(c,e,td),$ $ S_2'(c,f,td),$
$ \aux_2(c),$ $ S_1'(c,b,tss),$ $ S_2'(c,e,tss),$ $
S_2'(c,f,tss),$
$ R_1'(a,b,fa),$ $ \nit{diffchoice}(a,c,e),$  $ chosen(a,c,f)\}$ \\
$M_3$=&$\{R_1(a,b),$ $ S_1(c,b),$ $ S_2(c,e),$ $ S_2(c,f),$ $
R_1'(a,b,td),$ $ S_1'(c,b,td),$ $ S_2'(c,e,td),$ $ S_2'(c,f,td),$
$ $ $ \aux_2(c),$ $ S_1'(c,b,tss),$ $ S_2'(c,e,tss),$ $
S_2'(c,f,tss),$ $ R_1'(a,b,tss),$ $ chosen(a,c,e),$ $ $ $
\nit{diffchoice}(a,c,f),$ $ R_2'(a,e,ta),$ $ R_2'(a,e,tss)\} $ \\
$M_4$=&$\{R_1(a,b),$ $ S_1(c,b),$ $ S_2(c,e),$ $ S_2(c,f),$ $ $ $
R_1'(a,b,td),$ $ S_1'(c,b,td),$ $ S_2'(c,e,td),$ $ S_2'(c,f,td),$
$ \aux_2(c),$ $ S_1'(c,b,tss),$ $ S_2'(c,e,tss),$ $
S_2'(c,f,tss),$
$ R_1'(a,b,fa),$ $ chosen(a,c,e),$ $ \nit{diffchoice}(a,c,f)\} $,\\
\end{tabular}\\

\noindent which correspond to the following solutions (they  can
be obtained by selecting only the tuples with annotation
$t_{ss}$):~ $r^{M_1}=  \{S_1'(c,b),$ $ S_2'(c,e),$ $ S_2'(c,f),$ $
R_1'(a,b),$ $ R_2'(a,f)\} $, ~ $r^{M_2}=  \{ S_1'(c,b),$ $
S_2'(c,e),$ $ S_2'(c,f)\}$, ~  $r^{M_3}= \{S_1'(c,b),$ $
S_2'(c,e),$ $ S_2'(c,f),$ $ R_1'(a,b),$ $ R_2'(a,e)\} $,~ $r^{M_4}
= \{ S_1'(c,b),$ $ S_2'(c,e),$ $ S_2'(c,f)\}$.

\end{document}